# Highly connected – a recipe for success

**Krzysztof Suchecki, Andrea Scharnhorst**

The Virtual Knowledge Studio for the Humanities and Social Sciences (KNAW)

Cruquisweg 31, 1019 AT Amsterdam, The Netherlands

**Janusz A. Hołyst**

Faculty of Physics, Warsaw University of Technology, Koszykowa 75, 00-661 Warsaw, Poland

ABSTRACT

In this paper, we tackle the problem of innovation spreading from a modeling point of view. We consider a networked system of individuals, with a competition between two groups. We show its relation to the innovation spreading issues. We introduce an abstract model and show how it can be interpreted in this framework, as well as what conclusions we can draw form it. We further explain how model-derived conclusions can help to investigate the original problem, as well as other, similar problems.

The model is an agent-based model assuming simple binary attributes of those agents. It uses a majority dynamics (Ising model to be exact), meaning that individuals attempt to be similar to the majority of their peers, barring the occasional purely individual decisions that are modeled as random. We show that this simplistic model can be related to the decision-making during innovation adoption processes. The majority dynamics for the model mean that when a dominant attribute, representing an existing practice or solution, is already established, it will persists in the system. We show however, that in a two group competition, a smaller group that represents innovation users can still convince the larger group, if it has high self-support. We argue that this conclusion, while drawn from a simple model, can be applied to real cases of innovation spreading. We also show that the model could be interpreted in different ways, allowing different problems to profit from our conclusions.

## INTRODUCTION

The basic question for innovation research is how an innovation can emerge and survive. Classically one can look at growth curves of adopters (a global perspective from above) and at social networks of information transmission between adopters (an inside perspective) (Rogers 2003).

In innovation theories different classes of mathematical models have been use to formulate hypothesis of the dynamic mechanisms behind innovation spreading (e.g., Cointet and Roth 2007; Pyka and Scharnhorst 2009; Mahajan and Peterson 1985; Saviotti and Mani 1995). One popular model is that of epidemics spreading. This model describes the diffusion of a new idea, a new behavior or a new technology as an infection in a population where we can differentiate between infected, non-infected and "immune" agents. Another relevant approach concerns the role of fluctuations to design survival strategies for innovations. One of the authors has, together with other authors, developed a stochastic theory on the use of fluctuations in dynamically growing niches (Bruckner et al. 1996; Hartmann-Sonntag et. al 2009).

In all these models, the main question posed is what are the conditions for the innovation to survive the competition with established solutions and how it can spread, becoming a new established solution. One aspect of particular interest is how innovation can emerge and spread in hostile, competitive environment. Innovation always starts with a singular event – with one inventor and consequently with a small number of innovators in the stage of early adoption. Often it has been argued that a critical size is needed for an innovation to spreading out widely. Different rationale can be given for the existence of such a "critical size" (Heal 1994). Unforeseen events (modeled as fluctuations) can extinct adopters, so that a promising innovation vanishes with its possible few carriers. In classical growth models (Fischer-Pry) of logistic growth reasonable growth only occurs after the population of adopters has reached a critical size. In epidemic models, starting from very few agents makes failure and success almost random. All these models assume that in case of two competing alternatives eventually the growth rate determines the outcome of the competition. If the "new" is better it will survive. This has been also called "Darwinian selection". But this automatisms is obviously not always acting. In socio-technological systems dominant designs can emerge. Brian Arthur has proposed a model which shows that in presence of increasing returns (a specific form of network enhanced growth rate) a lock-in in a present form (technology in his case) happens (Arthur 1989). Qualitatively one could say that the existence of many other users is both needed (think in terms of infrastructures emerging around communities of users) and fosters the spreading of a technology. It has been shown that mathematically this can be mapped to the problem of two coexisting attractors in a space of possible states. As long as the system is the basin of attraction of the one state, alternative states cannot appear or disappear quickly as a fluctuation. It needs an outbreak of critical size to overcome the barrier between the two alternatives. This case has been labeled as hyperselection (Eigen and Schuster 1977, Eigen and Schuster 1978; Ebeling and Feistel 1982; Feistel and Ebeling 1988).

Since the nonlinear positive feedback has such a powerful impact, it is only natural that the questions about "critical size" and technological "lock-in" are regarded as important in the field of innovation theories (Witt 1997).

Thus the question how can this critical size be reached in a competitive environment is very important for innovation studies.

Niches (small, relatively separate groups of potential adopters) are relevant to foster a survival in a hostile environment. Niche creation can be seen as changing boundary conditions of the system (Bruckner et al. 1996). Other survival strategies operate on the level of local information exchange. In this paper we will look at the influence of information dissemination patterns among the involved agents – how the shape of communication patterns can influence the innovation spreading.

Our approach bases on a specific type of complex network model (Suchecki and Holyst 2006).We discuss how the mechanism of critical size (as a variable on the macro level) can be replaced by a mechanism of high connectivity (as a variable on the micro level). More particularly, we propose to use communities and their different connectivity as alternative possible strategy for the survival of an innovation. By linking global survival with local interaction mechanisms we immediately can place our model in the thought tradition of Rogers. Rogers connected diffusion curves (global variables on the group or type level) with social networks of information exchange (local mechanism on the level of the agents). The recently emerging branch of complex networks (Scharnhorst 2003) in statistical physics has systematically studies the effect of different topologies of networks on the spreading of innovation (Cointet and Roth 2007; Morone and Taylor 2004; Koenig et al. 2009) and our approach relies on this development.

We would like to stress that we present the model in a very general form. Although we started in the introduction from social-economic innovation theories, the social "system" we have in mind eventually can be a social community, a market, an industrial branch or a political system and the adopters can be individuals, companies, organizations or institutions.

First, we have to answer the question what is an innovation on such an abstract general level we are talking about. We define innovation as something new within the system, such as new technology, practice, solution, custom or even opinion, concept or ideology. The criteria for it being an innovation is that it must have not been present in the system before. It could have existed outside the system in question, such as for example steel tools being innovation for certain Australian tribes, despite being widely used in other parts of the world for a very long time (Rogers 2003) – if we only look at those tribes, steel tools are an innovation. We can basically distinguish two kind of innovations – those that succeed and spread throughout whole system and those that don't, remaining either in some niches or disappearing completely. The innovations that destabilize current system and cause system-wide change are called systemic innovations.

What happens with the innovation, whether it succeeds or not, depends heavily on adoption behavior. From a common sense observation of adoption behavior, the decision of an individual on adopting innovation is often influenced by other individuals. Contacts with successful adopters might influence the decision of the individual to adopt the innovation themselves. In addition, the efficiency or desirability of innovations quite often increases if others are also adopting (such as cell phone being "better" if your friends have ones too, or following a certain fashion or habit is "better" when your friends also do, instead of you being alone at that). Moreover, it is not only imitation as social phenomena which fosters group adaptation. Often innovations are connected with an infrastructure or are rooted with other cultural patterns (Nakicenovic 1991). The existence of a large network of gasoline suppliers supports one type of combustion engines and hampers alternative once – based on gas for instance. The coupling between a certain browser type (Internet explorer) with a certain operating systems (OS) created competitive advantages which had been later even negotiated legally (Windrum 2001). These "environmental" effects are often mirrored on the individual, social level. They are sometimes called "network" effects (Rogers 2003, chapter 8). If we apply this observation to the protypical/archetypical situation where two solutions compete (one being an innovation), we can assume – for the time being – that it is often advantageous for the single individuals to use the one that majority is using.

A similar situation can be found when we consider the opinion formation. Conformity is an essential behavior in society. Opinions of others strongly influence individual opinions. This can also be represented in an abstract way as going to conform with the opinion of the majority. This influence has been modeled before as "social impact" (Latane 1981) or desire to have neighbors similar to yourself (Schelling 1971).

In both the innovation spreading and the opinion formation the preference of individuals to make the same choice as the majority leads to the situation where small groups having different opinion or using an innovation are destined to disappear. In this paper we show that if the structure of the interactions between individuals is modular, meaning there are groups that are tightly connected inside, while weakly in-between, then it is possible for such minority groups to persist, and under certain conditions even spread their attributes (innovative practice, opinion, behavior or problem solution) throughout the whole system.

We discuss these effects with a model of stochastic dynamics on a network which can still be treated analytically. We also present a simulator tool which allows to play with rules and effects. The "Model Description" section carefully describes all aspects of the model we are using, focusing especially on the relation of the model with real world effects. The "Model Definition" section is more mathematically oriented and it defines the model in unambiguous way. "The System" section describes the environment – the set of agents interacting, especially the connections between these agents, that are responsible for interactions. The "Analytical Results" section introduces the results of analytical investigation into the model, along with its consequences. "The Simulator" section presents the simulation tool we have developed. Finally "Discussion" section explains how the findings concerning the model are relevant to real world and the "Conclusions" section gives few concluding remarks.

## MODEL DESCRIPTION – CONCEPTS AND SYSTEM BEHAVIOR

The model assumes there is a set of agents, each having a binary attribute. We shall label all agents by an index $i$, for further easy notation of specific agent's attributes as $s_i$. On the general level of mathematical structure, this attribute is a placeholder for a type of behavior, an opinion, an idea, a technology. Since it is binary, there are only two possible options to be chosen which is expressed by two possible values of the attribute. For the sake of being able to use a mathematical description we assume that those two values are -1 or +1. This is of course a strong restriction. It means that no in-between (half decided) behavior, no multiple choice and no choices on a larger set of option are modeled. However, such simplification captures the core issues of choice and difference itself, as well as allows analytical treatment of the model. Many models describing social phenomena use binary opinions/attributes in a similar way, such as the Schelling model (Schelling 1971) or models incorporating Latane social impact theory (Latane 1981; Lewenstein et al. 1992).

All agents interact with only a limited subset of other agents. Agents that interact are considered to be connected with each other, while those that do not interact are not connected. This allows us to treat agents as vertices and interactions as edges, and treat the whole system as a graph. We shall often refer to it as network (see figure 1). A subnetwork is a part of the whole network.

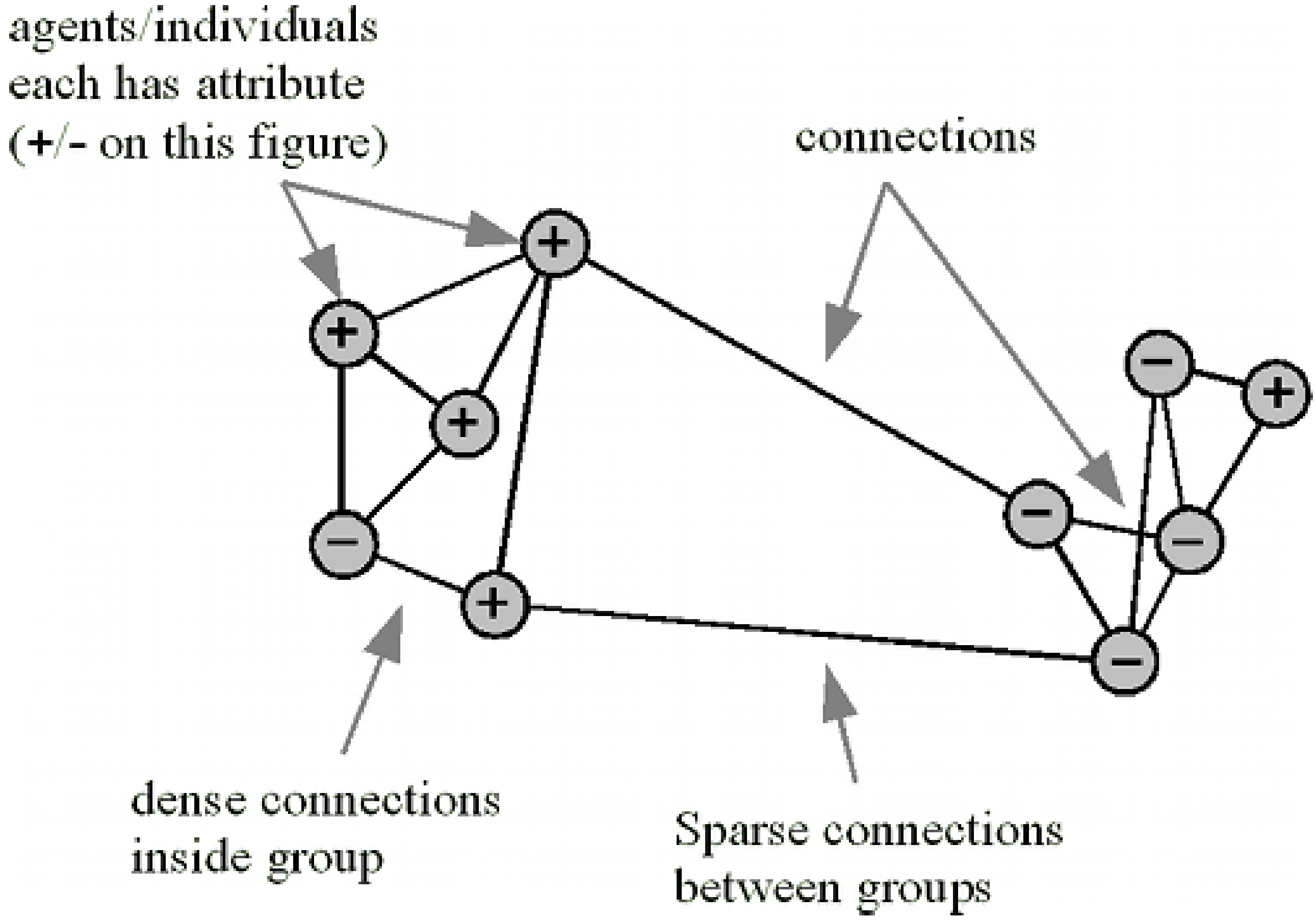

Figure 1. The illustration of the system, containing a set of agents each having binary attribute (represented as +/- in the figure, formally they are assigned values +1 and -1) and connections between those agents. The resulting network has a modular structure, with two densely connected groups and sparse connections in between them. The connections play a crucial role in the dynamics, as they define what agents interact with what agents during the attribute dynamics.

The nature of the interactions that is represented by edges is central for any social interpretation of the model. It is obvious that a single person cannot interact with everyone. Limited time and communication possibilities mean that each person can interact only with a limited number of other people. Interactions are possible by physical proximity or, more importantly, through various communication means (letters, phone calls, meetings) between people that know each other, or have some relation. The interactions we are talking about are essentially such potential channels of information exchange. If we interpret the model as opinion spreading, then edges would show mutual discussions and/or imitation behaviors that would result in one individual exerting influence over the opinion of the other and vice versa. In the diffusion of innovations, the edges may represent personal contacts between adopters, that are often decisive for the decisions on adoption innovation or not (Rogers 2003) – those that adopted would  influence others to do the same, while those that did not would influence others to do the same. It has been shown by Rogers, that the personal interactions and influences are crucial when deciding to try out an innovation. They are much important than non-personal communications such as commercials or information booklets. We believe that mediated communications and influences can also play a role here. The key here is trust. Even if communication is mediated, as long as it is personal – a person speaking, not hiding behind institutions or anonymity, then there is possibility for trust, that makes the communication influential. Thus while the connections represent mostly potential personal interactions, they may include also mediated ones. Still, all of them are personal – an agent is influenced by other agents, not mass-media or other public information channels.

In this paper, we keep the discussion on the abstract level and so we do not define the exact nature of interactions but keep them general and interpretable in many possible ways. It is however worth to note, that we consider the connections between agents to represent be static and potential interactions, not the actual interactions. Actual interactions or influences happen only along those connections. So far we defined interaction between two agents in binary way – it either exists or it doesn't. This can be extended and an edge attribute called weight can be introduced. It is a real number that indicates how strong the interaction between connected agents is. If the weight is zero, the interaction does not happen and the existence of edge has no impact on dynamics – it is as good as nonexistent. Thus, the introduction of edge weights gives a smooth transition between "interacting" and "not interacting". The introduction of edge weights allows to describe and understand networks in greater detail (Yook et al. 2001; Newman 2001), but on the other hand increases the complexity.

In our model we shall assume that the weight of all edges in the system is the same value $J$ and we shall call it interaction strength. We do not use heterogeneous weights, since the added complexity makes it harder to understand the causes of observed phenomena. For example, in our case, the phenomena observed in the model can be explained without relying on heterogeneous edge weights, what means they are not responsible for its emergence. The interaction strength $J$ can be treated either as edge weights or as a general system parameter, and is required in the mathematical formulation of the model. We choose to use it as edge weight, because it allows to understand easier what this parameter is responsible for, as well as allows for easy extension of the model by applying heterogeneous weights - this is possible within the mathematical formulation of the model, but would invalidate analytical results obtained that assume constant $J$.

In summary, the system is a network, composed of a number of agents identified by their index $i$, that possess a binary attribute $s_i$. The edges in the network represent interactions between agents.

The model describes the dynamics of the agent attributes. The network structure (connections between agents) can change, but only as a result of external intervention, not as a part of internal dynamics. The dynamics causes only changes in the attribute values of agents, while the interaction strength and network topology remains constant. In other words we consider a dynamics taking place on a stable network topology. The attribute changes are results of actual interactions, while network connections represents the potential, static interactions.

Agents change their attributes with time, based on the local rules. This means that an agent's changes of attribute are only influenced by his neighbors – agents that he is connected with. Unconnected agents have no direct influence on the attribute changes. These local rules can differ ,such as agents trying to match majority of neighbors in Ising model (Stauffer 2008; Galam 1997), or adopt attribute value of one random neighbor in voter model (Dornic et al. 2001), and describe different ways the topology of the network is relevant for individual decision making. For example in the Schelling model (Schelling 1971) the attributes describe agent's race: white or black and the local rules dictate that if your neighborhood is in majority different from you, you will exchange attribute values with a different agent in the neighborhood, similarly dissatisfied with its neighborhood (exchanging attributes is equivalent of people exchanging the place they live in and therefore the place in the network of physical proximity). In the voter model an agent unconditionally adopts the attribute value of one randomly selected neighbor, representing an opinion change due to interaction (connections represent your contacts, but they are actually "used" only sometimes and at random).

Regardless of the wide range of possible rules that can be implemented in modeling, the nature of agents and attributes, the nature of connections, the nature of dynamics and the dynamical rules all have to be considered together, not separately. We cannot investigate smoking behaviors, interpreting agent as smoking or not smoking, depending on the attribute, while interpreting connections as e-mail exchange patterns and applying majority dynamics to this all. Unless we could prove that a person decides to smoke or not smoke, based on what majority of the people he corresponds with through e-mail does. We cannot interpret the smoking/nonsmoking behaviors as "social behavior" separately, e-mail exchange as "social relations" network, while majority dynamics as "social pressure" separately, claiming that the model shows impact of social pressure on the behavior patterns.

It is worth to note that in the most general case the dynamics of the attributes may also depend on some additional properties of the agent. For example the agents with many connections might act differently than those with few (Snijders et al. 2009). But in our model we disregard that possibility. Such different behaviors are very hard to describe in consistent, mathematical way. All the diversity of agents rests within the topology of the network in this model. This is of course a simplification, meaning that our model cannot explain any effects that arise from the heterogeneity of agents other than topology-related. It is important to remind however, that models are always simplifications.
The basic rule of dynamics is to change an attribute to match the attribute of the majority of neighbors. So an agent with attribute -1 would change to +1 if he had 4 neighbors, 3 of which would have attribute +1. If he already had an attribute matching that of a majority, then he would not change. Since the attribute changes depend on the state of the majority, we say that this model implements a majority dynamics. There are multiple ways for a model to implement a majority dynamics, such as using majority rule (Krapivsky and Redner 2003) or through agent preferences as in Schelling model (Schelling 1971). The dynamics of the latter are actually very similar to our model, which is equivalent to Ising model. A relation between these two was discussed by Stauffer (Stauffer and Solomon 2007)
Note that agents in this model have no memory – only the current situation influences an agent behavior, not anything that happened in the past, even agent's own past.

It is important to note that the majority dynamics is not strict, meaning the agents not always conform with the majority of their neighbors. There is a parameter we call individuality, that controls how often the agents will conform and how often they will not. While we call it individuality, it is only one possible interpretation. The parameter comes from statistical physics, where it is known as temperature. It can be given different interpretations, such as individuality (personal differences), tolerance (as seen from the systemic perspective) (Mimkes 2006), resistance, rebellions, non-bidable behavior or non-obedience against general rules. The meaning of this specific dynamic characteristics depends , depending on how the rest of the model elements (agents, connections, attributes) are interpreted.
It might sound puzzling that in the model The individuality is defined as a global parameter. The parameter sets a probabilistic level on which an agent will disregard the majority rule and act differently. For each concrete event in the simulation of the model, the agent is selected randomly and the final decision is based on probabilities. In these random elements which cause fluctuations around deterministic rules singular events, personality, the un-foreseen and un-foreseeable are hidden. In the case that the agent does not conform, their individual choice is represented as random, since individual or occasional circumstances cannot be described in terms of trends or universal laws (Eigen and Winkler 1993).
When the parameter of individuality is zero, the agents always conform with the majority of their neighbors. As it increases, agents increasingly often act individually, deciding to change or not to change at random. When the individuality is infinite, the agents never pay attention to what neighbors think and choose their attribute at random (although random choice will coincide with the majority about half of the time). The majority dynamics can be understood as a social pressure to adopt certain behavior (Latane 1981). The decision of the agent will depend on the relative values of social pressure given agent feels and the level of individuality within the system. The higher the individuality, the higher social pressure must be to convince the agent to behave in the same fashion. Let's consider a specific agent. If there are 30 agents in the neighborhood, all having the same attribute +1, then it requires a very high individuality for the agent to disregard the majority. On the other hand, if 16 agents have attribute +1 and 14 have opposite, then even relatively low individuality can cause nonconformist behavior, since the majority is very "weak". If the agent is not convinced to follow the majority, then his behavior is modeled as random, as the background of such individual decisions are too complex and beyond the scope of the model.

Although we explore the model in the context of social application its roots are in physics and it is known there as the Ising model. In physics, the attributes and agents are called spins and the individuality is direct equivalent of temperature.
The majority dynamics present in the model are actually derived from the energy changes and statistical mechanics. The Ising model has been created to describe the properties of magnetics – why and how they generate inherent magnetic fields. It is important that also in this application it significantly simplifies reality. However, it captures the essential property of the system – emergence of spontaneous ordering through very simple majority dynamics. Although developed originally to describe magnetization, the Ising model can be treated as a very generic model thanks to its simplicity. Its mathematical formulation can be adapted to different phenomena. Other examples of such generic models are differential equations of Lotka-Volterra-Type (Peschel and Mende 1986) or skew distributions (Mandelbrot 1982). The motivation to use it for a social phenomena description is similar – we hope to explain collective behaviors through very simple rules, so that a relation between agent-level behavior and emerging system-wide phenomena can be established.

## MODEL DEFINITION – MATHEMATICAL FORMULATION

The mathematical formulation of the model is as follows. Each agent i (i={1,N}) has an attribute $s_i \in \{-1,+1\}$ and $k_i$ neighbors belonging to his neighborhood $K_i$. The individuality in the system is labeled $T$ (for continuity with the roots of the model), while interaction strength (edge weights) equals *J*.

The dynamics is asynchronous, meaning that the agents in the network are subject to the dynamics at random, one by one. The other possibility – synchronous dynamics means that all agents are subject to dynamics at the same moment. While synchronous dynamics may seem natural, it is not so because both time and attribute are discrete values in the simulation. If synchronous dynamics is applied artifacts can emerge – phenomena arising purely from the simulation procedures (in this case oscillatory behaviors). Asynchronous dynamics do not suffer from such problems (Lawson and Park 2000).

In our model, we define the flow of time by discrete time steps. Each single time step each agent should be updated once. However, we use an implementation that only statistically fulfills the rule. We consider one time step to be equivalent of *N* single agent updates, where *N* is the total number of agents. Since the choices are random, in a single time step an agent can be subjected to dynamics more than once, while some may not be subjected at all. However on average, each agent is subjected once per time step.

The probability for an agent i to assume the state $s_i$=+1 equals

$$P(s_i = +1) = \frac{1}{1 + \exp\left(-2Jh_i/T\right)}$$

where $h_i$ is the sum of neighbor attributes – so called local field

$$h_i = \sum_{j \in K_i} s_j$$

The above equation exactly defines how individuality, social pressure (local field) and chance to conform are related to each other. It is worth to note that this equation does not work for $T$=0, since it would produce a division by zero. However, the limit for $T \to 0$ coincides with strict majority dynamics (except in case of $h_i$=0, when the behavior is completely random).

Both individuality *T* and interaction strength *J* are parameters of the system, so the same global values are used for all agents. Both parameters appear in the equation together, as *J/T*. This means, that only this ratio is important for the dynamics, not the exact values of each. It follows, that both can be combined into a single parameter. We reduce them to a relative individuality $T'=T/J$ and use it in place of "pure" individuality. The relative value shows the impact individuality has on the dynamics for any *T* and *J*.

This feature of the model and can be understood as follows: a strong individuality in an environment with strong influences produces the same result as a weak individuality in an environment of weak influence.
In summary we have a majority dynamics model, where each agent $i$ has an attribute $s_i$, and uses a majority dynamical rule, with agent individualism measured by $T$ and interaction strengths $J$.

## THE SYSTEM – WHERE THE DYNAMICS TAKE PLACE

Our study focuses on a specific network topology. We investigate the model on a modular network, that consists of two coupled subnetworks. The concept of modularity or community structure is important in social systems. The community as a whole is not homogenous. Groups of similar individuals often interact with each other more often than with individuals that are different. The community has a structure, where smaller, tightly connected communities are sparsely connected into a larger whole (de Nooy 2009).

We assume that the whole system consists of two (sub)networks and a number of edges that interconnect them. The most natural way to visualize such two subnetworks is to see two separate groups of agents. The visualization used later for the simulation tools presents the two networks separate from each other. In this visualization the number of links between the network parts can easy be followed. However, it is important to realize that we could also visualize them as center and periphery or as two intertwined networks. Both subnetworks are part of one and the same network. They can be only distinguished by the topology of the connections. It is worth to note that in case the connection density between subnetworks is as high as inside subnetworks, then they are no longer distinguishable and the division is arbitrary and artificial. In fact, as the number of interconnections increase, the networks are increasingly difficult to distinguish. When we create the coupled networks, we first create the subnetworks separately, so all agents clearly belong to one or the other. However, when interlinks are introduced in large number, this becomes unclear. Some agents may end up with more links to the second network, than the one they initially were in. Thus, when we only look at end effect (and use community detection algorithms), they rather belong to the second network, not the first. Such occurrences are rare when the number of interlinks is small, but increase quickly when the number of interlinks approach the critical value where networks are indistinguishable. As the networks become more and more undistinguishable, the analytic predictions we make become gradually less certain.

The internal structures of the subnetworks do not have to be fixed for our model to work. However, we have made some assumptions about that structures to actually perform analytical and numerical analysis.

For the analytical part, we have assumed that the subnetworks are random. In this case, random networks do not simply mean Erdos-Renyi random graph, but all network topologies that have random connections.

The Erdos-Renyi random graph model is a simple model, where an actual realization of the network possesses a finite fraction of all possible edges in the network chosen at random. If there are $N$ vertices, then there are $N(N-1)/2$ possible connections (the number of pairs of vertices) between them. Out of them a fraction $p$ is actually present in a network. In practice, any possible connection will exist with a probability $p$. This is the basic random network model.

In general, random networks may be created in many different ways (such as the Barabasi-Albert (Barabasi and Albert 1999) model or the Watts-Strogatz model (Watts and Strogatz 1998) for $p$=1), but share a common property. The connections between agents are not correlated with each other. Square lattice (where agents are positioned in a regular pattern and connected by regular connections, forming square lattice) for example has very correlated connections. By knowing small part of the whole network, we can infer the rest of the structure and connections. For example, seeing three connections between vertices of one of the "squares" in such lattice, we immediately know, thanks to the correlations, that a fourth connection also exist between the vertices, that closes the loop. Moreover, we know that there are two additional connections for each of the vertices (for a total of four). This is true with 100% probability, thus showing that the correlations are very strong. If the lattice was diluted (some connections removed at random), or if some of the connections had randomized ends, then we could not tell with certainty, showing that the correlations are weaker. In contrast, knowing small part of random networks gives us absolutely no information about the parts of the network we don't see. The connections could be there or not and we have no way of telling what the probability is.

It is worth to note that assuming our networks, that are models of real-life communities, as simple random networks is a reductionist approach. We disregard many of the features of real systems. However, this is still much better model than regular lattices. In real life, people or institutions have contacts and communications with many others. But these relations never take shape of a regular lattice. They are much, much more complex. The random network assumption captures this feature, although it dispenses with the complexity, simply modeling it as randomness. While never as correlated as lattices, real social networks still possess significant correlations (Newman 2002). However, capturing all features of real systems is still beyond current state of research and still an open question. Moreover, such highly complex systems would prove immune to analytical investigations and make drawing correct conclusions much harder. Thus, we settle with a simple random networks, to focus on what can be already be observed in such simplified systems.
A feature of random networks that we use in the analysis is that the probability of connection existing between any two vertices $i$ and $j$ is proportional to a product of their degrees $k_i$ and $k_j$. It is actually true only for uncorrelated random networks, but in this paper we are only considering really uncorrelated random networks or those with correlations Since we are using vertex degrees, we already look at existing network, not at a creation process. The property is easy to understand. As both vertices have more connections (to different other agents), the chance that one of them is actually the one between them increases. . Note that without knowing global properties (such as total number of connections and vertices), we still know nothing about probability of the connection existing, because the product $k_ik_j$ still needs to be divided by total number of connections to get actual probability. Thus the property of not being able to tell whether connection exists or not still holds. It is worth to note that being able to use probabilities instead of actual connections is important feature of random networks. We are actually investigating a whole class of networks at once, not a specific network. This is possible because the connections in random networks can be treated statistically. Thanks to investigating a whole class at once, our results are very general and valid for the whole class. In contrast, focusing on specific topology and then broadening claims to a whole class of network topologies carries danger, that the effects we observed are exceptional and abnormal for the class as a whole, not typical. Thus, the conclusions we can make have a very general applications.

For the numerical analysis, we have to choose a more specific topology, since we need to actually create these networks. We have assumed there, that subnetworks have the internal structures of the Barabasi-Albert scale-free networks (Barabasi and Albert 1999). This theoretical model is based on principles of evolution of network and preferential attachment and produces a cohesive (without disconnected parts) scale-free network as a result. We have chosen it as a representative of scale-free networks.
The Barabasi-Albert model is a specific model of an evolving network. It starts with a fully connected cluster of $m$ agents. At each time step a new agent is added and it creates $m$ edges connecting to already existing agents. The agents to connect to are chosen preferentially, meaning at random, but with probability to connect to given agent $i$ being proportional to degree $k_i$ of that agent. This means that a Matthew principle ("rich gets richer") (Merton 1968; Merton 1988) is present and as a result a power law in the degree distribution appears, with scaling exponent $\gamma=3$. As said before, in our model we do not concern ourselves with the evolution of such network. We use the evolution only as a way to obtain a scale-free network as the place where the dynamics run. We consider the network to be static for the purpose of modeling dynamics.

In our system, we have two subnetworks A and B, containing $N_A$ and $N_B$ agents. The parameter $m$ that determines the number of connections each new agent makes to the existing network during its evolution may be different for each and equals $m_A$ and $m_B$. This differentiation allows us to describe a situation where two groups of different size and different internal cohesion exist.

Our two subnetworks A and B are interconnected by $E$ connections. There are several ways the connections can be introduced. The simplest way is to randomly choose agents from both networks to connect. However, if the probability to connect agents is constant in a network, then all agents will tend to possess statistically the same number of interconnections. This means that agents with few internal connections have disproportionate large number of interconnections. That is why we choose the agents to connect preferentially. The probability to choose an agent for interconnection is proportional to its internal degree. In this fashion we obtain a modular network, where each agent has statistically the same fraction of the interconnections. We designate this fractions as $p_A$ and $p_B$ for both networks respectively (if the networks have different sizes and densities, then $p_A$ and $p_B$ must be different since the number of interconnections is the same on both ends).
This assumption about how the links are introduced has two reasons. First, it has social explanation – people with small number of acquaintances are less likely to get new ones. In reality this might be related to different optimal number of acquaintances for each person. It also models the desire of everyone to prefer to know and acquaint popular people, rather than some unknown ones. The second reason for such assumption is that it greatly simplifies the analytic calculations.

In summary our systems consist of two connected subnetworks. We have subnetwork A, that is a scale-free network consisting of $N_A$ agents and having average degree $<k_i>=2m_A$. The second subnetwork B is described by similar set of parameters $N_B$ and $m_B$. The subnetworks are connected by $E$ edges that connect preferentially chosen agents in both networks.

ANALYTICAL RESULTS

Our analysis focuses on the problem of two subnetworks having opposite starting attributes. This can be understood as two separate groups, each holding different opinion or using different practice or solution. The interesting scenario is when these two groups come into contact with each other and start interacting. Such a clash between the two groups can be represented as two subnetworks starting forming connections between themselves. The added interaction results in a competition between the two initial attributes.

Let us first focus on more basic behavior of the model. The model described so far can be treated analytically to determine some principal behavior. It was found out that the system has three different stable states, depending on the parameters. Stable states are states to which a dynamic system approaches and then persists in them. The stable state may be a stationary configuration of agent's attributes, but dynamical stable states are also possible. Repetitive patterns of agent attributes, or completely random behaviors for all agents can also be considered stable states, provided they persist. For example, a complete ordering of all agent's attributes – global consensus, is a stable state for the model, provided that relative individuality is minimal or zero. On the other hand, if relative individuality is very high, the stable state is completely random, which statistically correspond to relatively stable half/half distribution of attributes (even though individual agent attributes are constantly changing) for large system. Stable states correspond to attractors in theory of complex dynamic systems. In general stable states persist in the system until an external influence changes the structure of interactions in the system, so that new stable states emerge for these new conditions. Since the system may have more than one stable state, the achieved one is determined by the initial conditions. Noise, which represents either external or internal random disturbances, may switch the system from a stable state to another, but usually the waiting time for such event to occur would be long to extremely long compared to the timescale in which systems approach a certain stable state.

Since our model is basically identical with Ising model, we know that a single network (of almost any topology) has two stable states: *ordered* and *disordered*. An *ordered* state occurs when one attribute is dominant – most agents share the same attribute and the shared attribute does not change. The most *ordered* possible state is consensus where all agents have same attribute. An *ordered* state can have some agents which have different attributes than the most, but they are merely fluctuations that appear from time to time. *Ordered* states are stable when the relative individuality is low. For our model there are actually two possible *ordered* states, with most attributes +1 or most -1. Since the model is symmetric, both states behave exactly the same and are essentially identical. Therefore, we can often talk about just "*ordered* state" without referring specifically to either one. It could be compared to saying that a person with a pending decision is either "decided" or "undecided". If the exact decision made is not important for what we are investigating, then the "decided" state can be treated as just one state, ignoring the fact that on more fine level many different decisions may have been taken. Mathematically, it is the case with our model, and we can talk about just one *ordered* state. Still, the difference between these two might be important in practical interpretations, even though they play no role for the dynamics and general investigation.

Aside from *ordered* state, the Ising model can be in a *disordered* state. It occurs when no attribute is dominant in the network and most agents change it fast and randomly. *Disordered* states are stable when the relative individuality is high.

The stable states can be summarized in a table:

| Internal state | System state |
|---|---|
| *ordered* (+1) | *ordered* |
| *disordered* | *disordered* |
| *ordered* (-1) | *ordered* |

Table 1. Possible system states for single network

In the case of connected coupled networks, the situation is much more complex and there are three stable states: *disagreement*, *agreement* and *anarchy*. It can be summarized as before in a table:

| Internal state A | Internal state B | System state |
|---|---|---|
| *ordered* (+) | *ordered* (+) | *agreement* |
| *ordered* (+) | *ordered* (-) | *disagreement* |
| *ordered* (-) | *ordered* (+) | *disagreement* |
| *ordered* (-) | *ordered* (-) | *agreement* |
| *disordered* | *ordered* (+) | not stable |
| *disordered* | *ordered* (-) | not stable |
| *ordered* (+) | *disordered* | not stable |
| *ordered* (-) | *disordered* | not stable |
| *disordered* | *disordered* | *anarchy* |

Table 2. Possible system states for two coupled networks A and B.

Out of 9 possible combinations of stable states of subnetworks, 5 of them are stable states for the system, while 4 of them are not stable. *Disagreement* occurs when both subnetworks are *ordered*, but they have different dominant attributes (corresponds to -1/+1 or -1/+1 internal orderings). *Agreement* occurs when both subnetworks are *ordered* and share the same attribute (corresponds to

+1/+1 or -1/-1 internal orderings). *Anarchy* state means that both subnetworks are *disordered*. The combinations of *order-disorder* are not possible, because the internally *ordered* network imposes its own order onto the second network, causing it to be *ordered* too. It could happen in case where the connection between both network is almost or completely nonexistent, but then both subnetworks could be considered separate and not as belonging to a single system for practical purposes.

Table 2 shows all possible combinations when we assume that two subnetworks can clearly be identified. We have argued earlier that the transition between modular networks and one networks is continuous. For a pair of fixed subnetworks ($N_A$, $N_B$, $m_A$, $m_B$ are fixed), there are two parameters that influence which stable states can emerge: relative individuality $T'$ and number of interconnections between the networks $E$. In an analysis of the model each stable state can be calculated depending on the values of these two parameters. If plotted in the parameter space (T', E) the different nature of the states can be indicated as different regions (called phases) of that space, separated by lines (phase boundaries). This graphical representation is called a phase diagram. The phase diagram allows to determine the possible outcome for each model setting (in terms of parameter values). Figure 2 shows the phase diagram of the model when the two sub-networks have same size and density. The boundaries in this phase diagram indicate critical values of relative individuality and have been calculated (Suchecki and Holyst 2006; Suchecki and Holyst 2009, Suchecki 2008).

To interpret the phase diagram one has to take into account that it does not represent the trajectories – the dynamical changes of actual system states. It only shows what stable states exist for given parameter values. Since the model is stochastic, it is potentially possible for the system to appear in any state for any parameter combination. However, some states are much more likely to appear than others. For example in the phase of *anarchy*, it is technically possible for a system to appear *ordered* for a moment, but the chance for such an occurrence is very, very low – for example, for a small system of 100 agents, the chance for system to appear fully *ordered* for a single moment are about 1 in $10^{30}$, what means that if we observe system that does 1 million time steps per second, we would have to wait many times longer than the age of the universe to observe single such event. Due to these probabilities, that are extreme in some cases, we can say that for a given parameter combination, the system is or will be soon in one of stable states and be quite sure about it.

We can also consider a system "moving" through the phase space – this means that its parameters would be changing during the dynamics. As long as such system remains in one phase, nothing special happens, but should it cross the phase boundary, the system state will change to conform with possible stable state in the new phase. The very important point here is that the speed at which parameters change is relatively slow compared to the speed of dynamics. The whole stable state reasoning and analysis is only valid if the parameters are quasi-stationary and don't change nearly as fast as the dynamics itself proceeds. Investigating systems where parameters change with speeds similar to the rate of dynamics is much more complex problem and related to nonequilibrium statistical mechanics. We do not attempt to describe this kind of systems here and assume that parameters change slowly and the dynamics has enough time to reach stable state each time the parameters are changed..

The phase diagram (Figure 2) shows that *disagreement* can exist only if both relative individuality and inter-network interactions are  low. The maximum relative individuality $T'_{c1}$ depends on inter-network interaction strength *E*. The higher the interaction strength, the lower is the maximal still tolerable relative individuality $T'_{c1}$. Since at low relative individuality level the fluctuations are low, this allows subnetworks to persist in opposite *ordered* states when combined with weak inter-network interaction.

*Agreement* can exist in a broader range of relative individuality and for any inter-network interaction strength. The maximum relative individuality for the *agreement* state is noted as $T'_c$ (in Ising model it is known as critical temperature). It is worth to note that in the *disagreement* phase, both *disagreement* and *agreement* can exist – both are stable states for these parameter values. Which one is actually the state the system is in depends on its history. If the system started out or entered this phase being in the state of *disagreement*, it will remain in *disagreement*. If it started out in *agreement*, it will remain in *agreement*. Additionally, due to stochastic nature of the model it is possible for the noise to change the system from *disagreement* to *agreement*. It is theoretically possible for reverse to be also true, but it is extremely unlikely. The chance for such transition is relatively low and the larger the system (looking at number of agents) and further away from *disagreement*/*agreement* phase boundary, the smaller the chance is.

Above a critical level of relative individuality $T'_c$ only the state of *anarchy* is stable. No order prevails in such conditions – the fluctuations destroy any ordered clusters that might emerge.

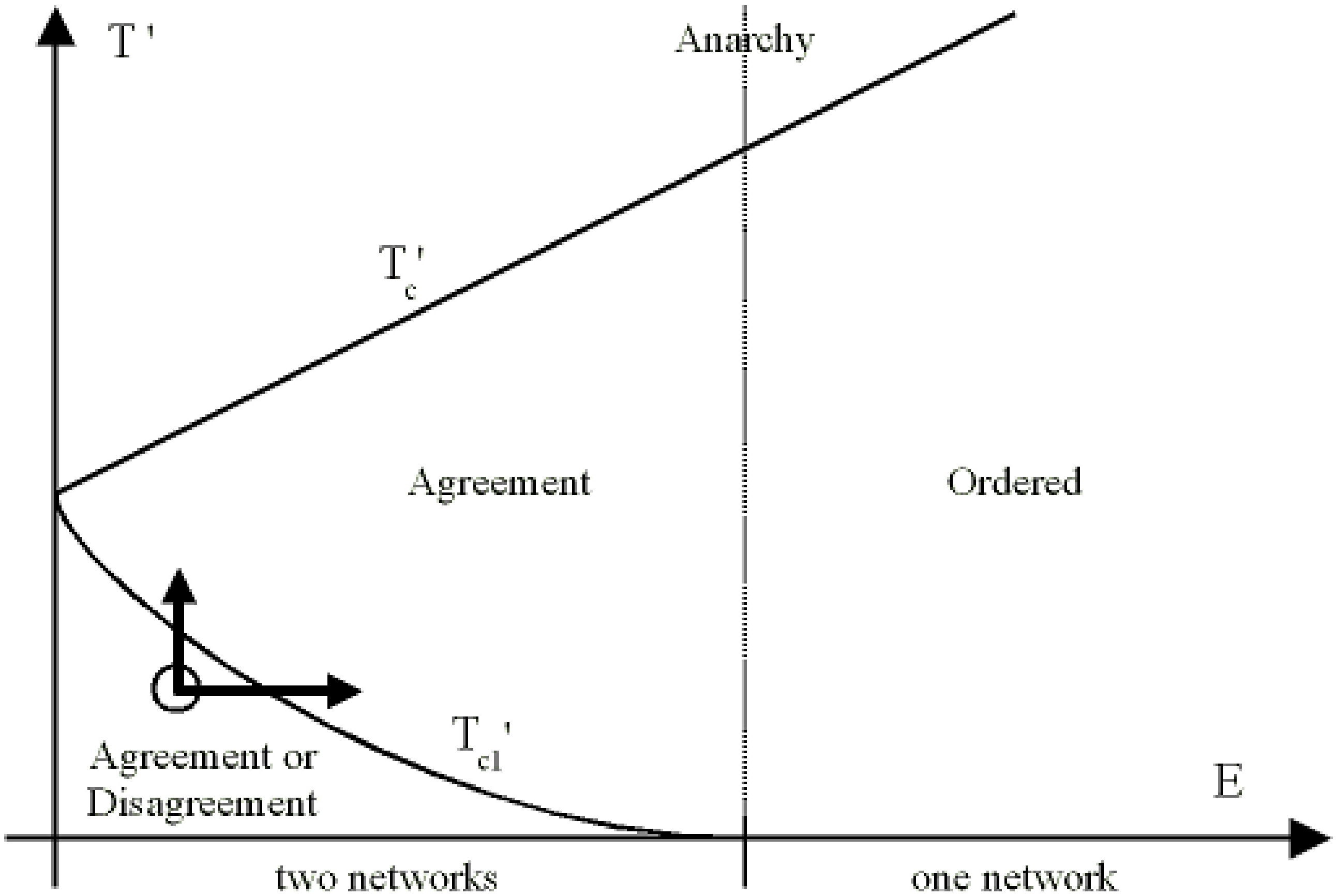

Figure 2. The phase diagram of the majority model (Ising) on two coupled random (Barabasi-Albert) networks. Three possible stable states: *disagreement*, *agreement* or *anarchy* can exist, depending on the relative individuality (T') and the number of edges between networks (E). There is a threshold along the E axis, above which the networks are unseparable and therefore can be only treated as single network – therefore there is single network *ordered* state instead of coupled network *agreement*. If a system starts out from *disagreement* in the point marked with a circle, then either increasing individuality or number of edges (changing parameters along the indicated arrows) results in arriving in *agreement*-only phase. This forces system to leave *disagreement* and reach *agreement*. Note that in the lower-left phase both *agreement* and *disagreement* are stable states. The system can exist in either one, depending on initial conditions. If the system enters this phase (or starts out) in *disagreement*, it will remain in *disagreement*. If it enters in *agreement*, it will remain in *agreement*.

Let us now consider the system of two *ordered* networks with opposite dominant attributes (state of *disagreement*). If the relative individuality and interconnection number are low, the state can persist for a long time. However, if the relative individuality or number of interconnections increase enough, the system finds itself in a place where only *agreement* is stable. The state of the system changes rapidly (compared to speed of parameter change) from *disagreement* to *agreement*. However as it has been mentioned before, there are two possible outcomes. The system can order with +1 being dominant attribute value, or it can be -1.

This can be interpreted as a process, where two social groups that have different opinions, values, procedures or technical solutions are brought into contact, either by outside force, or some inside interactions not covered in the model. If the considered opinion or other attribute is subject to social pressure with a majority-type dynamics, then one of the opinions become dominant throughout both groups – a result of influences between the groups. This happens if the groups interact strongly enough (there are enough connections), or if the relative individuality and unpredictability is too high (the individuals do not stick enough with their own group enough when it comes to opinion). If the interaction is weak and relative individuality is low, the system will persist in state of *disagreement* for a long time (although it is possible that one of attributes become dominant through fluctuations).

In case of innovation spreading, the two networks can be interpreted as communities of individuals who already adopted the innovation and those who did not. The adopters will usually convince others to adopt, while non-adopters will remain mostly distrustful and discourage others to adopt. This is mirrored in the majority dynamics of our model. The starting situation of community of adopters and community of people who didn't adopt is equivalent to the system in the state of *disagreement*. If the networks interact strongly enough, and relative individuality is high enough, then the system finds itself in the *agreement* phase and the actual system state will change to *agreement*. The possible outcomes are either adopters convincing non-adopters to adopt, or for the non-adopters to convince others to discontinue innovation use. Of course the interesting issue is whether the innovation will turn out successful (adopters convincing nonadopters) or unsuccessful (when nonadopters convince initial adopters to discontinue the use of innovation).

The phase diagram we have shown earlier (Figure 2) is not the only possible phase diagram for our model. Depending on what we consider a parameter, we could have different phase diagrams. For example, if instead of relative individuality T' and number of interconnections E, we consider relative individuality T' and number of agents in subnetwork B $N_B$, then our diagram will look differently (Figure 3). We have chosen these parameters for our example, since it allows us to use the phase diagram for additional explanations in the numeric results section.

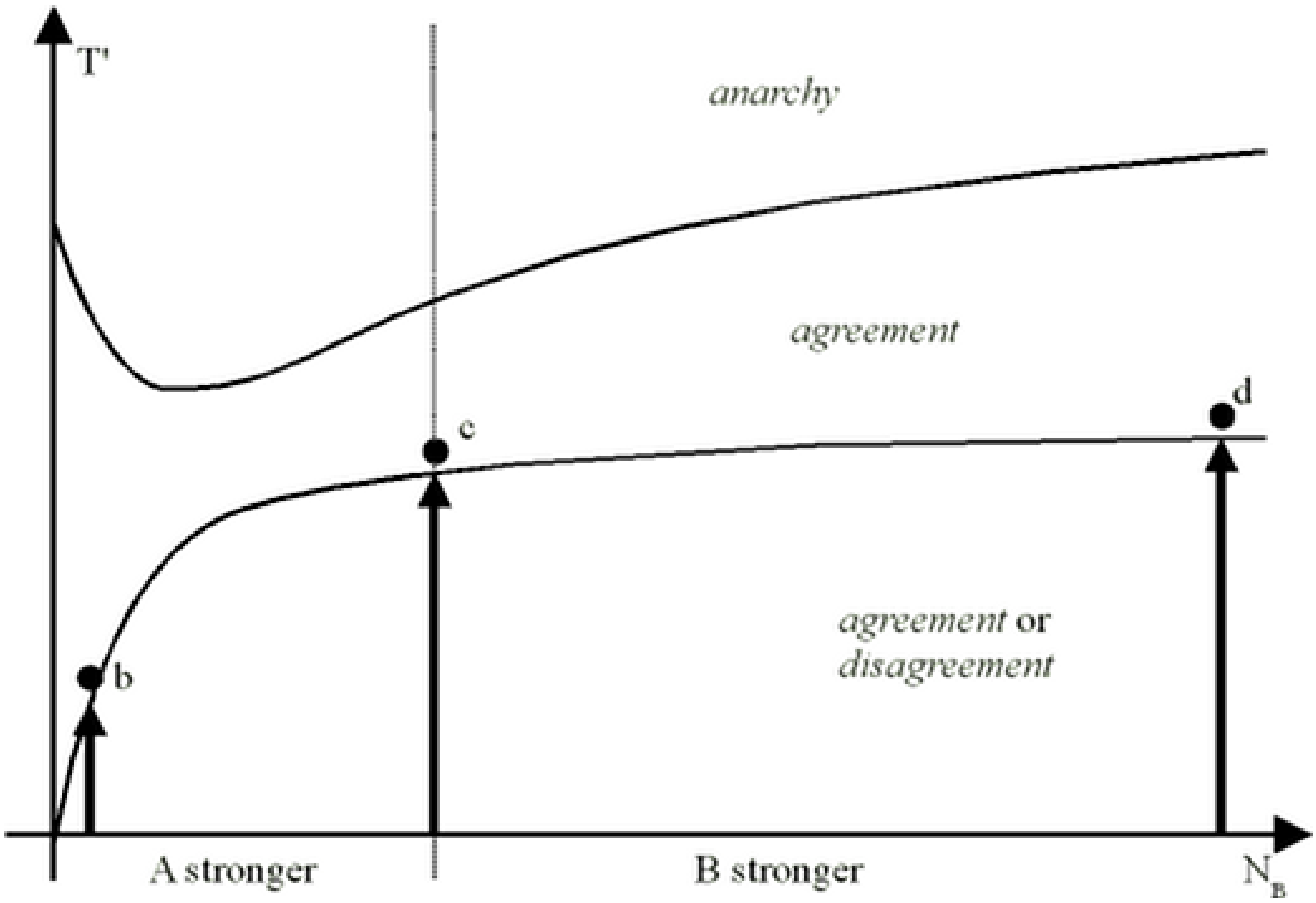

Figure 3. The phase diagram of the majority model (Ising) on two coupled random (Barabasi-Albert) networks with relative individuality (T') and size of one of subnetworks ($N_B$) as parameters. Like the phase diagram presented in Figure 2, it still shows the same three possible states: *disagreement, agreement* and *anarchy*, just in different parameter space. There are three points marked b,c and d, that qualitatively show the points in phase space, where the simulation resulted in graphs b-d in Figure 4.

In the case of a simple model like ours, it is possible to predict who will be the winner statistically. Due to random nature of the both the dynamics itself and the network structure, it is impossible to make a prediction that is certain, but it is possible to say victory of which network is more probable.

It turns out, that there is a certain measure of the subnetwork that decides how likely is it to win. We call this network indicator "strength". The stronger of the two subnetworks wins more often. This "strength" can be calculated mathematically (Suchecki and Holyst 2006) and expressed as $<k^2>/<k>$. The $k$ are agent degrees in given subnetwork, while the brackets mean an average over all agents in given subnetwork. For simple random networks, where all agents have similar degree, this value is basically almost equal to just average connectivity $<k>$, but the presence of highly connected hubs can significantly raise it above that value. For the Barabasi-Albert network it has been theoretically calculated that

$$\frac{\langle k^2 \rangle}{\langle k \rangle} \approx \langle k \rangle / 4 \cdot \ln N$$

where N is the number of agents. While being qualitatively true, it fails to predict this value precisely in practice. However, the qualitative conclusions stand valid. $\langle k^2 \rangle / \langle k \rangle$ for Barabasi-Albert network does depend linearly on connectivity $\langle k \rangle$ and logarithmically on size N, just like the equation predicts . This dependence of the "strength" on the connections density means that the cohesion and self-support of the group are most important, while size is of lesser importance (due to logarithmic dependence). For a network of agents with similar degrees, the size proves to be outright irrelevant - $\langle k^2 \rangle / \langle k \rangle$ does not depend on the size.

### *Numeric results*

So far we have discussed possible stable states of the system depending on parameters of the dynamics and the role of network topology in the competition between the sub-networks. For each point in the phase space – each parameter combination – we can observe the temporal evolution of the system into this stable state using simulations. Depending on the initial conditions (the network topology) we observe different time lines. Some characteristic runs are presented in  Figure 4.

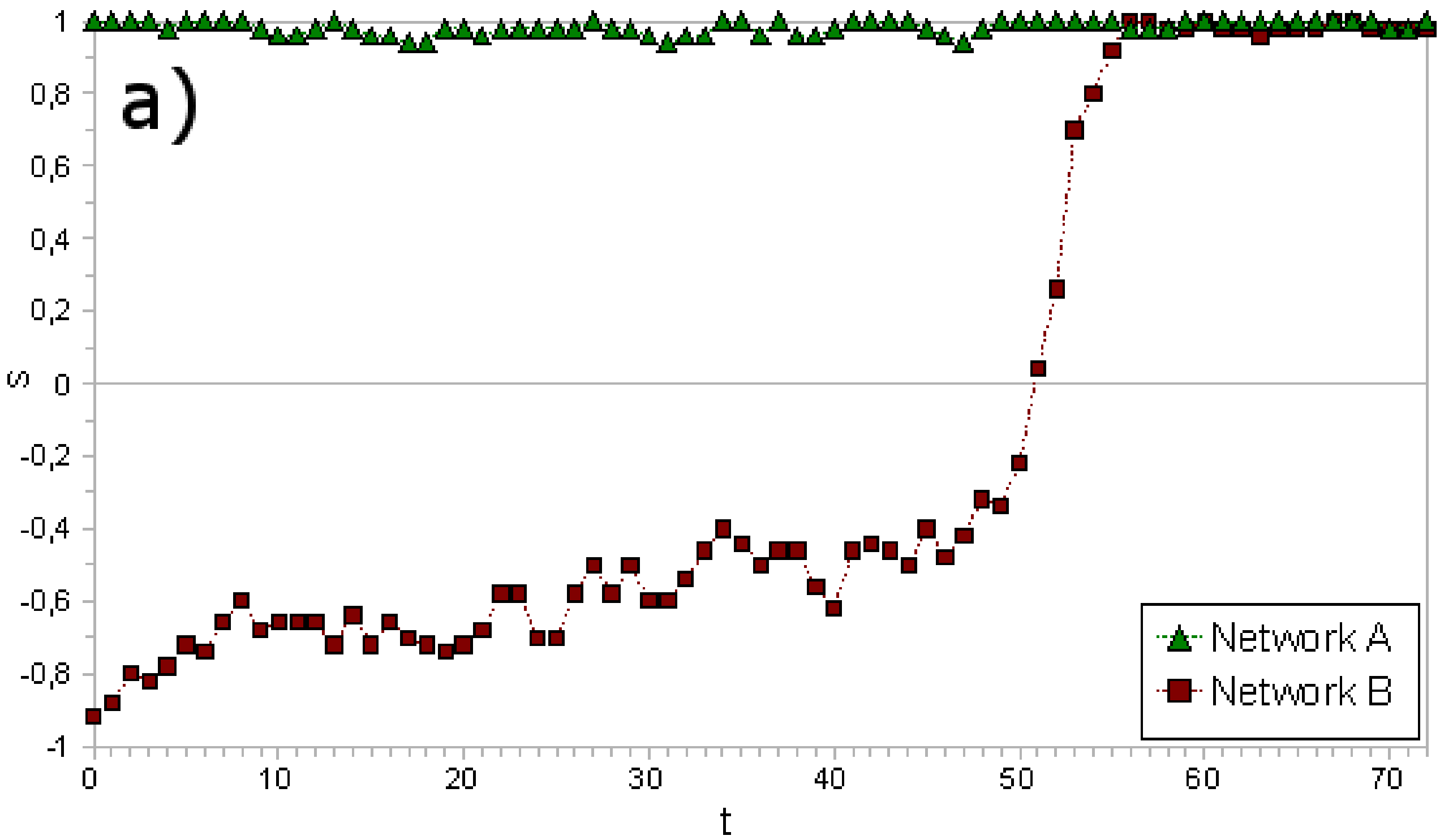

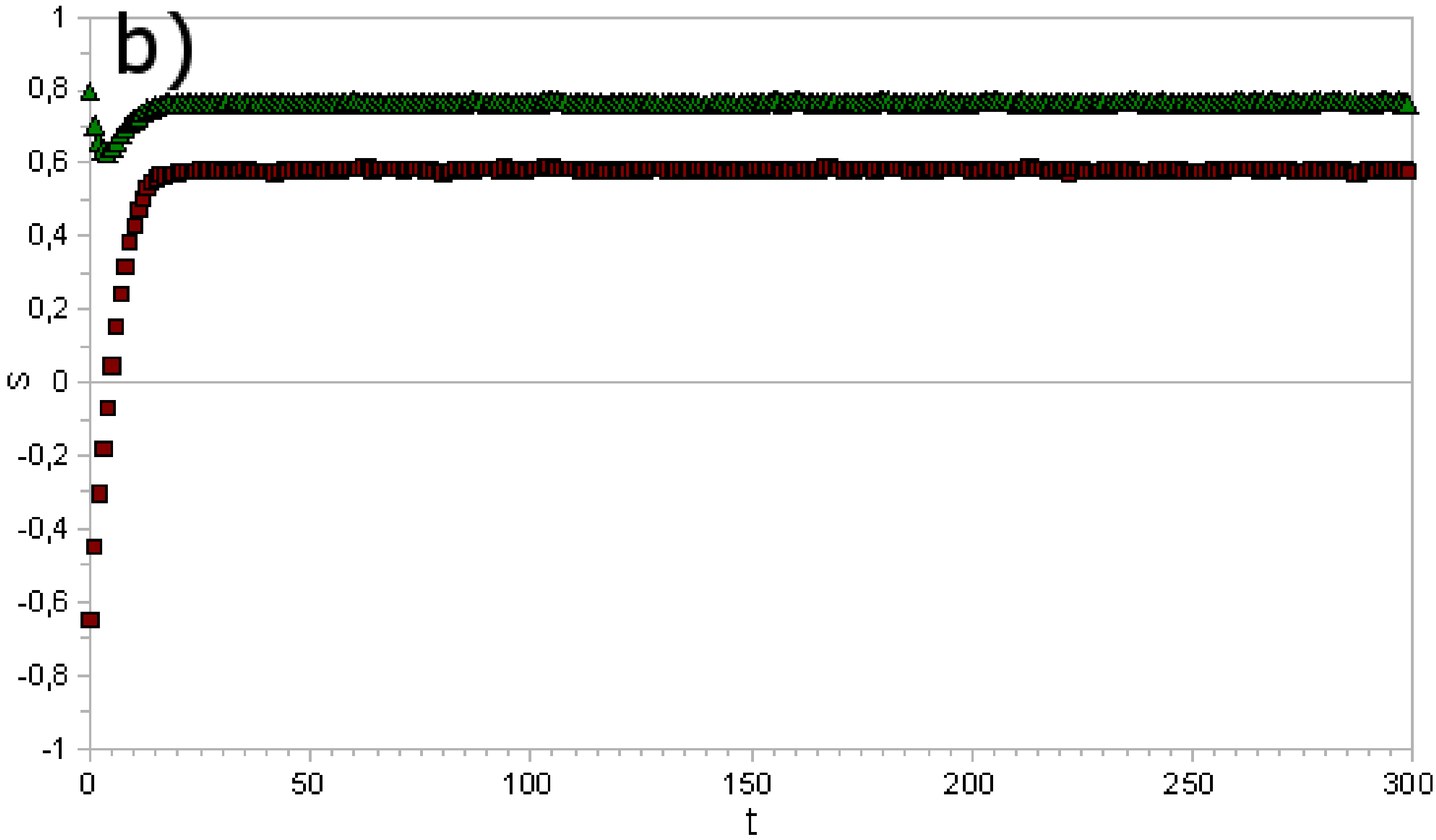
b)
s
t
1
0,8
0,6
0,4
0,2
0
-0,2
-0,4
-0,6
-0,8
-1
0
50
100
150
200
250
300

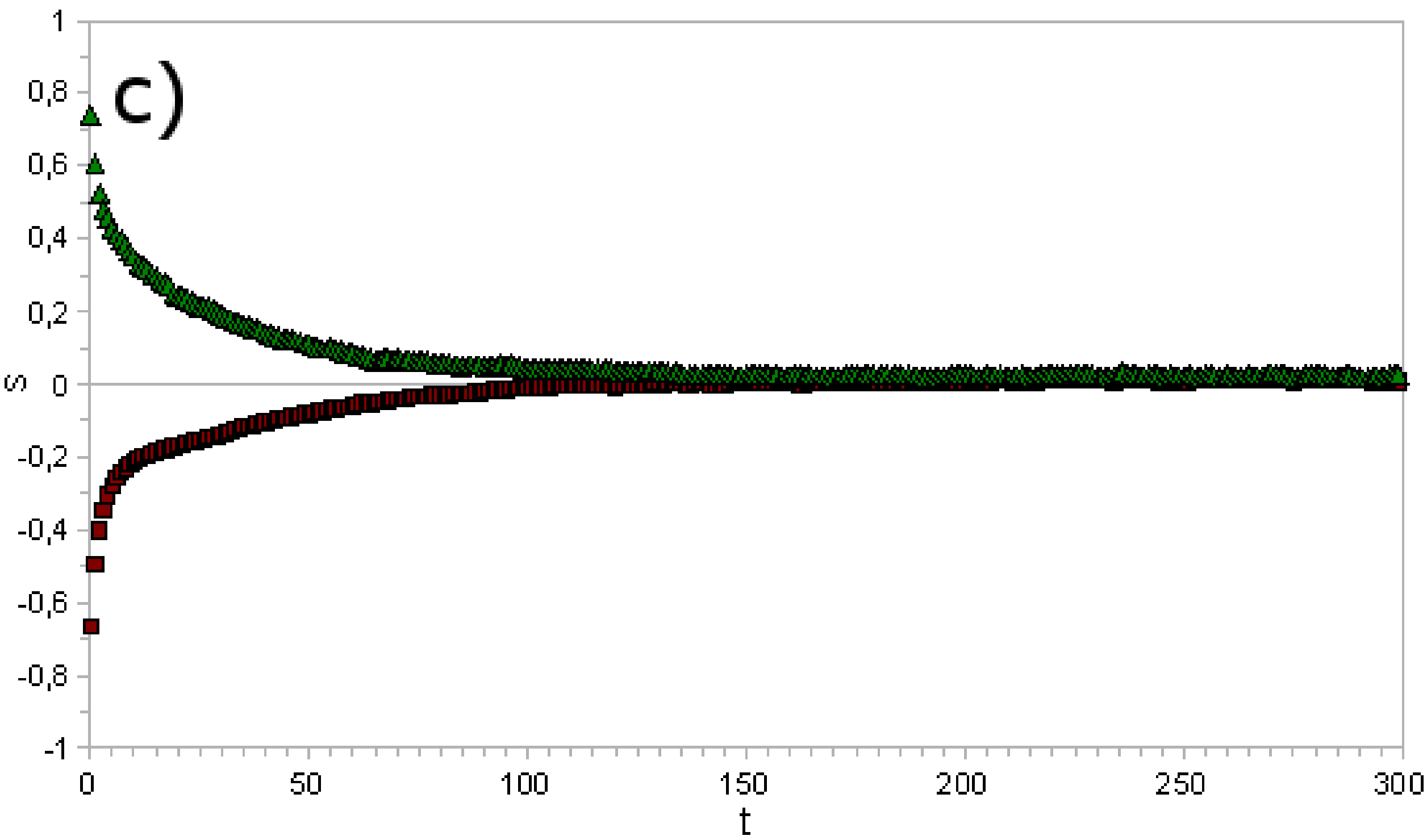
c)
s
t
1
0,8
0,6
0,4
0,2
0
-0,2
-0,4
-0,6
-0,8
-1
0
50
100
150
200
250
300

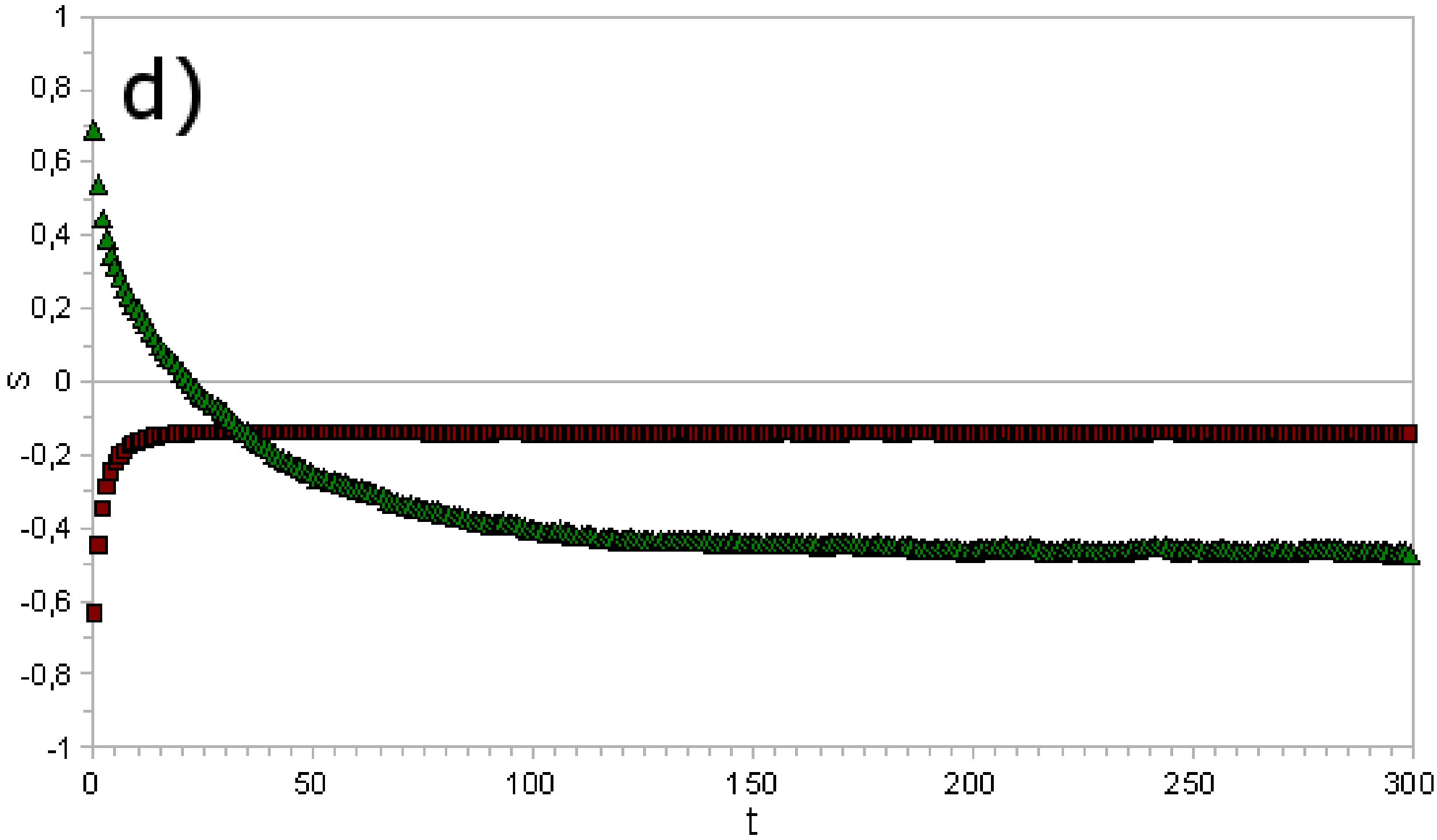

Figure 4. Plots of average attribute values *s* against time *t* in two subnetworks (A and B) in the model simulation. The plot a) shows single, typical simulation starting from *disagreement* state, while plots b)-d) show averages over 1000 simulations. In all cases subnetwork A had $N_A$=100 agents and average degree $<k_A>$=6, while subnetwork B was of different sizes and smaller average degree $<k_B>$=4. There are E=100 interlinks between the subnetworks and the relative individuality was chosen so that the system is forced into *agreement* state. a) $N_B$=100, T'=4, b) $N_B$=100, T'=5, c) $N_B$=702, d) $N_B$=10000, T'=6. Note that while both $N_B$ and T' change between cases, graphs b-d all are situated in almost the same point in the phase space – with T' just above the *disagreement-agreement* phase boundary. Changes to $N_B$ change phase diagram, so T' has to be changed too to place the system in similar situation.

For each network we calculate an average attribute value. Plot a on Figure 4 shows how the average attribute values for both subnetworks change in time. At the beginning, when all agents within a subnetwork share the same attribute value, the average is equal to that value (-1 or +1). This is our usual initial condition. The average value will change during dynamics, almost always straying from extreme initial values of +1 or -1. This is because when relative individuality is non-zero, there are practically always some agents who temporarily have different attribute than the most.
The average attribute value lines are closely tied to fraction of agents with given attributes. After all, the average attribute <s> depends on numbers of agents with attribute +1 and -1 directly: $<s>=N_{+1}/N-N_{-1}/N$, where $N_{+1}$ and $N_{-1}$ are number of agents with attribute +1 and -1. The average however, contains all information that both $N_{+1}$ and $N_{-1}$ contain, because the total number of agents in our model do not change, so lines for both would be forced to mirror each other.
On the plot a in Figure 4, the subnetwork A "wins", because around time t=50 the subnetwork B

changes its average attribute from initial negative to positive. Thus both networks have positive attribute values and a state of *agreement* is reached, with subnetwork A as “winner” (since it's the initial attribute of A that now is dominant throughout the whole system).
As mentioned earlier, we can predict which network will win only statistically. Because of that, observing single runs like the one presented in plot a is not very useful. To have some statistical credibility, we would need to observe thousands of plots and compare which subnetwork won each time. It is a lot easier to delegate that work to computer, and instead of single run, plot an average over several runs. This average is created by adding up the average attribute graphs created in multiple simulation runs. Let us look at Figure 4, plots b-d. These present the averages over 1000 separate runs. Each time, the individual runs resembled somehow plot a – they had considerable fluctuations, and at one point one of the subnetworks switched to the opposite state, thus the system reached *agreement* state. However, it was not always the same network that “won”. Let us look at plot b. In this case network A is the usual winner. Adding many runs where network A had positive average attribute created significant, positive average value, seen on the graph. The fluctuations are gone – they were random, so over 1000 different runs, they cancelled out themselves. This is important feature of averages over realizations, that the fluctuations disappear. However, there were few individual runs where network A actually “lost”. However, since there were many more times it has “won”, the average turned out strongly positive. Looking at the line for network B (usual “loser”), we can see it is initially at -1 (since in every single run the starting condition was the same, the average has to have exactly this value), but then quickly changes to positive. This shows that around time t=20, in most individual runs the network B already “lost” and had positive average attribute.
Now let us take a look at graph d. In this one, network B is the usual “winner”, and we can see behaviors very similar to the ones in graph b – the average for usual “winner” (network B) remains with its initial sign (negative in this case), while the other network (network A) crosses the zero and changes its average attribute to match that of the “winner”. We can see that the line for network A changes much slower than the line for network B in graph b. This means that it usually took more time for network A to “lose” - around time t=20 the line is still around zero, which means that during half the individual runs it was positive (still didn't “lose”) and during half it was already negative (already “lost”). The graph d features another difference from graph b – the lines cross each other. This actually has little practical meaning and is only related to the actual values of average attributes both networks reach.
Now let us look at graph c. Here, we see averages over different runs go both to zero. What does this mean ? This can actually have two different interpretations. One is that both networks become *disordered* and therefore their attribute values both go to zero. But looking at individual realizations we know it is not true. The second interpretation is that subnetworks were *ordered*, but about half of the time it happened with positive attribute value, while about half the time it happened with negative. After adding it up all runs, the net result is zero. This is exactly the case with this graph. Both subnetworks are *ordered* (after reaching *agreement* state), but the attribute – positive or negative is random, with same chances. This means that there is no usual "winner" here – subnetworks "win" about half the time and "lose" about half the time. At the very beginning neither network had yet the chance to win, so both have attribute values close to their starting values. As the time passes, the lines represent the outcome of the "competition" more and more, while representing initial conditions less and less. This is true for all graphs, and that's why the lines don't reach the final value immediately, but take some time.

Let us now discuss the parameter values that led to the results presented in graphs b-d. The subnetwork A is always the same, with $N_A$=100 agents and average number of neighbors $<k_A>$=6. The subnetwork B has average number of neighbors $<k_B>$=4, but the size differs from plot to plot: b) $N_B$=100 agents, c) $N_B$=702 agents, d) $N_B$=10000 agents. The sizes have been chosen so that in case b subnetwork A has higher "strength" (9,42 versus 7,55), both subnetworks have about the same "strength" in case c (9,87 versus 9,95), and in case d network B is "stronger" (10,13 versus 13,06). The relative individuality T' is also varied in these three cases. This was done to actually place the system into a very similar state, not to make it different. Let's look at the phase diagram at Figure 3. The three black dots qualitatively represent the parameters ($N_B$, T') we used for the simulation when creating graphs b-d. All of them are in similar place – just above the *disagreement–agreement* phase boundary. The value of T' in each case was chosen to place the system just there. This way, despite having different sizes, we are qualitatively in the same situation in each case we compare.

Since our initial conditions are fully ordered *disagreement* in each case, we can treat the system as if the starting relative individuality T' is equal to zero (the only value where the system is completely ordered, with no fluctuations). Now, when starting simulation we were setting the value of relative individuality T' to some non-zero value. This is equivalent of moving the system in parameter space – from initial T'=0 to a certain positive T'. Since the system crosses *disagreement–agreement* phase boundary, it will change its state from initial *disagreement* to *agreement*. Thus one we consider of the networks a "winner" when its initial attribute dominates whole system. As can be seen, on all plots b-d the stronger subnetwork is the usual "winner", as explained above. This confirms the statement that this strength statistically determines the winning network. It is one of the more important results presented in this paper. It is worth to note that only slightly better connectivity of network A (average number of neighbors is 6, while in network B it's 4) makes network A stronger unless the opposition is seven times as large (702 versus 100).

Thanks to using formal description of the model based on statistical mechanics, now we can make an additional prediction on the matter of one or the other network “winning”.
In real life situations, it is often hard or downright impossible to determine the degrees of individual agents. Questions arise when a given relationship should be represented as an edge and when it is too weak or does not contribute to the attribute changes of an agent. While it is extremely hard task to measure the actual interactions, it is usually a lot easier to measure the results of these interactions meaning the attributes (opinions, innovation use). For situations where this model may be applied, attributes are binary – such as a yes/no opinion or use/non-use of innovation. Measurements of these attributes are much more plausible, and in fact in certain situations they are already conducted (pre-election surveys or the elections themselves).
Here is where the statistical mechanics come in handy. It turns out, that by looking at the history of the measurements of average attributes, we can actually predict the outcome of the “competition” between the subnetworks. If we observe the average attributes in both our subnetworks, we may point out which one is more likely to win. The subnetwork with larger fluctuations of the average attribute value is the one that has smaller “strength” and therefore is the one more likely to lose. Let us refer to Figure 4, plot a again. Note that the subnetwork that experienced larger fluctuations was the one to “lose”. This is typical behavior, although due to random nature of the model it is always possible that the outcome will be opposite. It is inevitable that one can only operate on probabilities when dealing with stochastic systems.
In short, looking at the history of the real system, and especially on the stability of the attributes within subgroups, we can predict which subgroup will emerge victorious if they start interacting more strongly, or if more noise, fluctuations and individualism is introduced.

## THE SIMULATOR

While the model itself is simple, the emerging behavior isn't so obvious. It is useful when the dynamics can be visualized, and that one can observe all the dynamics and their results himself. A simulation software for the model that has the visualization integrated is therefore required. We have decided to create a new program, as none of the existing programs or software packets could be easily adapted to perform as we wanted. There was already a similar tool developed before, but it had limitations in the user interface and still used physics names and labels. We elected to build upon the idea of this simulator, developing a tool which is more suitable for non-physicists.
Our simulator is striving for following goals:

- Allows simulation of majority dynamics model on coupled networks
- Intuitive interface
- Descriptions and labels suited for broad audience

The simulator has been programmed in Java language using JUNG (Java Universal Network/Graph Framework) library. It allows the creation of single or coupled networks with a specified parameters: number of agents $N_A$, $N_B$, density of connections $m_A$, $m_B$ and number of inter-network connections $E$.
Additionally it is possible to add agents and connections by hand or load network from a file (Pajek Net and GraphML formats are supported). This allows potentially any network topology desired. The network structure can be modified before or during the dynamics simulation. Agents can be formally assigned to one of two groups (all agents in subnetworks are automatically assigned to two groups when coupled networks are created).

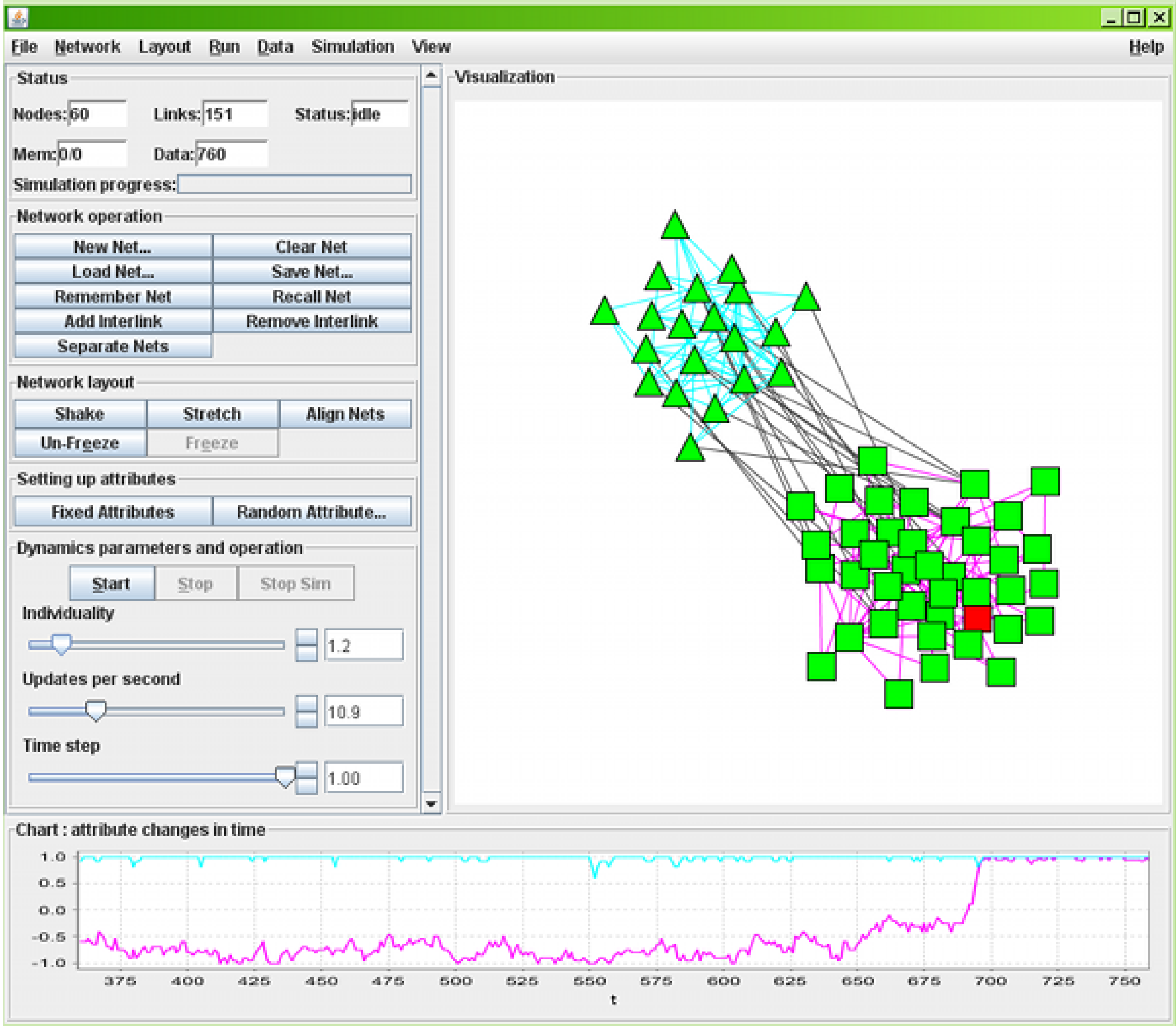

Figure 5. Screenshot of the simulator. The main part is the visualization of the two groups of agents, distinguished by their different shapes and color of intra-group links. To the left is the control panel that allows to construct the network and control the simulation parameters. On the bottom, the chart representing the history of changes of average attribute values in both groups is situated.

The network is visualized as a set of agents represented by circles (unassigned agents), triangles (agents in group A) and squares (agents in group B). The agents have a color representing their attribute – green represents value +1, while red represents value -1. Connections between agents are visualized as simple straight lines. The color of connections depend whether the connection is between agents in the same group (cyan for group A, pink for group B) or not (black).

The agents are distributed in the visualization window randomly at first, but the connections act like a springs, changing layout dynamically so that connected agents remain relatively close to each other, allowing to see structure easier (this can be turned on and off).

The simulation uses the Ising majority dynamics and asynchronous update method to avoid artificial oscillatory behaviors as described earlier in the paper in the Model Definition section. It uses the exact formula for chance of attribute being equal to +1 after agent update described in Model Description section.
The dynamics can be started and stopped as desired. During the dynamics, the history of average attribute changes in two groups and several other parameters can be recorded in the graph and displayed during the dynamics, so the user can track changes in accordance with what happens in the visualization. The resulting data can be viewed and exported. It is also possible to perform simple statistical simulation of the system, where the visualization and graph are disabled for maximum performance.

The simulator (including the code) is available and can be found at <http://simshelf2.virtualknowledgestudio.nl/activities/duonet-simulation-tool>

## DISCUSSION

The aim of the paper is to point out a specific phenomena – the competition between different sub-networks in a networked system. It discusses the existence of the stable *disagreement* state and the transition from *disagreement* to *agreement* depending of network topologies and a specific dynamics running across the networks. Specific focus is given to the possibility of a smaller group "winning" against a larger group, as natural assumption would be for the larger group to "win". The idea of smaller community "winning" against a larger one might be directly related to innovation diffusion and opinion formations. It shows, that the topology of the relations between individuals or adopters are enough to allow minority spreading from small, close-knit niches outwards to the whole system.

The model presented is a theoretical, abstract model. It features only one agent behavior – conformity with the neighbors (although limited). Because of its abstraction it can be related to several very different real-world situations, although it cannot be said that the model represents either of them. We will discuss two such situations, one being the diffusion of innovations, the second being opinion formation.

In the field of diffusion of innovations, the innovation spreading proceeds through several stages:

Knowledge, Persuasion, Decision, Implementation and Confirmation (Rogers 2003) (in the original Rogers' book of 1962, the stages were slightly different: Awareness, Interest, Evaluation, Trial and Adoption). Out of these stages, the Persuasion and Decision stages are most dependent on external influences. During that time adopters gather information from trusted sources, quite often friends or professional acquaintances. The evaluation of the innovation, and therefore decision to adopt or not depends on the information gathered and influence being subjected in this stage. Our abstract model, being a model of interactions between individuals, is applicable to these stages, where the interpersonal communications and dynamics play role. For other adoption stages, the decisions are mainly based on individual evaluation of innovation. This is not covered by the model at all.
Various other innovation models also exist, more than a few sharing a common feature – increasing returns for innovation adoption. This means that the more people adopt, the more profitable it becomes to adopt too. This leads to a "lock-in" situation where it is impossible for a new, emerging innovation to establish itself and spread throughout the system. Yet as we know new innovations constantly emerge. The question is, how do these innovations fare at the beginning ? How do they survive the competition with existing, omnipresent solutions ?
The results of this model may shed a new light on this problem. The innovations may be firstly adopted only within a close-knit niches. Then, due to interactions with other communities, they might spread, just like the attributes of the smaller group spread throughout the system in our model. The relevance of niches for innovation spreading is very well known (Saviotti and Mani 1995), but the crucial point is how the innovation can leave the niche. The interesting point is that our model does not assume one attribute to be "better" than the other, meaning that even innovations that are marginally better or not better at all than existing solutions might fight their way through the market, provided they are backed up by a strong group (in our model – highly self-supportive group with dense connections). Also, as we know from lock-in phenomena, simply "being better" might be of absolute no help in some situations. In an earlier paper one of the authors has shown that via a dynamic growing niche a still "better" innovation can even overcome an hyperselection or lock-in. In a recent paper (Scharnhorst et al. 2009) one of the author has discussed different other possibilities to overcome hyperselection which all finally make use of the role of fluctuations when trying to tunnel through attractor basins. The present network model points to an additional survival strategy. Being highly connected and this way protected against the hostile environment while at the same time active in attacking this environment seems to be a reasonable strategy as well. The "strengths" of a network is like an invisible castle around a subgroup in a community. A subgroup which may win even when smaller. If this network property is mathematical equivalent to potential adoption rates of higher order remains on open question for further research.

The second area where the results of our model may apply is opinion formation. The social behavior exhibits conformity, since it is essential for coherent communities. This conformity is what our majority dynamics represents. Non-conformal behavior is included in the model as random decision making.

In this case, the overtaking of a smaller group over the larger may show how the minority opinions may spread and eventually become one of the major opinions in the system.

It is important however, to realize that in both described situations, as well as in any others the model might apply to, the model itself is a theoretical abstraction. It offers no specific predictions or explanations. Its purpose is to show how a smaller, yet more organized group can win with a larger one, and show that such process does not require agents in both groups to be different. They can act in the same way, the only difference being how they interact between themselves. The model might offer an explanation also for the longevity and influence of so-called secret societies (Erickson 1981).

The results obtained from the model may help us understand the real mechanisms behind the innovation spreading, but they do not offer explanation themselves.

It is important to point out that the two presented cases are not the only ones where the model might be applicable. Due to its abstract nature, the model is very open to various interpretations. What we call agents, links and individuality might be interpreted in completely different manner. Examples might include agents being scientists and attributes corresponding to specific scientific issue or organizational solution, agents being companies and attributes corresponding to market practices, agents being internet servers while attributes corresponding to the default transmission protocols.

## CONCLUSIONS

We show, that in an abstract model related to innovation spreading and opinion formation, a possibility exist for a minority opinion or innovation to spread throughout the system. The model is simple binary state majority dynamics with a noise representing individuality.

The minority can "win" and spread throughout the system, despite majority dynamics. We show that the topology of relations between the individuals is enough to allow persisting minority, and that if the minority group has strong internal ties, it can "win" in the end, even against overwhelming majority. Heterogeneity of agents, or complex rules of choice are not required for such behavior. If the network of relations has a community structure, with groups of agents strongly tied together then if is possible for minority to “win”, provided that the minority group has stronger internal cohesion (stronger or more connections) inside that the majority group.